\documentclass[12pt,preprint]{aastex}
\ifx\undefined\iint
\newcommand{\iint}{\int\!\!\!\!\int}
\fi

\ifx\undefined\bmath
\newcommand{\bmath}[1]{\mbox{\boldmath$#1$}}
\fi
\ifx\undefined\mathbfss
\newcommand{\mathbfss}[1]{\mbox{\bf\textsf{#1}}}
\fi
\ifx\undefined\bm
\newcommand{\bm}[1]{\bmath{#1}}
\else
\renewcommand{\bm}[1]{\bmath{#1}}
\fi
\begin{document}
\title{
Continuous Limit of Multiple Lens Effect and the Optical Scalar Equation
}
\author{Hiroshi Yoshida}
\affil{Department of Physics, Fukushima Medical University,
Fukushima-City 960-1295,Japan}
\email{yoshidah@fmu.ac.jp}
\author{Kouji Nakamura}
\affil{Department of Astronomical Science,
the Graduate University of the Advanced Studies, Osawa, Mitaka 181-8588, Japan}
\email{kouchan@th.nao.ac.jp}
\and
\author{Minoru Omote}
\affil{Department of Physics, Keio University, Hiyoshi, Yokohama
223-8521, Japan}
\email{omote@phys-h.keio.ac.jp}
\begin{abstract}
 We study the continuous limit of the multiple gravitational lensing
 theory  based on the thin lens approximation. Under the approximation,
 we  define a new, light-path dependent angular diameter distance  
 $\tilde{d}$ and show that it satisfies the optical scalar equation.
 The distance provides relations between quantities used in the
 gravitational lensing theory (the convergence-, the shear- and 
 the twist-term) and those used in the scalar optics theory 
(the rates of expansion, shear and rotation). 
\end{abstract}
\section{Introduction}
The distance-redshift relation is one of the most important and
difficult subjects in observational cosmology (here, ``distance'' means
the angular diameter distance). Recent observations on Type Ia
supernovae with high-redshifts suggest the 
large deacceleration parameter $q_0=\Omega_0/2-\lambda_0$
\citep[$\Omega_0, \lambda_0$ are the present values 
of the density parameter and of the cosmological constant,
respectively;][]{SNIa}. 
The reliability of the suggestion depends on that of the distance 
measure to supernova adopted in the observation.
\par%
Since our universe is assumed to be described by the
Friedmann-Lema\^{i}re universe which is homogeneous and isotropic in a 
large scale, we have used the Mattig formula as the distance measure
\citep[e.g.][and see also Appendix A1]{Wein72}. However we know that
the universe is not homogeneous in a small scale and that light ray
propagating in the locally inhomogeneous universe is gravitationally affected by clumps
intervening between the source and the observer which is known as the
gravitational lens effect. Then we have to
estimate the distance to a source based on the light ray propagating in
the homogeneous universe in a global scale but locally inhomogeneous
universe. Unfortunately, however, we have not known the solution of the 
Einstein field equations describing such a universe yet. 
\par
A recipe to obtain a distance-redshift relation without the exact
solution of the field equations was presented by \cite{DR72,DR73}.
They gave a distance measure based on light bundle propagating away from all
the clumps between the source and the observer, where the rate of shear
in the scalar optics theory is negligible (the shear-free assumption). In
general, their distance measure is larger than that given by the Mattig
formula. Consequently, it yields the dimming effect on 
the observed flux compared with one based on the Mattig distance measure.
They also showed that the distance measure depends on the dimming effect
more seriously than the the gravitational magnification effect in the
Swiss-cheese model \citep{Bertotti66,DR74}.
\par
However the shear-free assumption does not always hold for light bundles
propagating in the general space-time. In fact \cite{WS90} showed that
when the light ray experiences multiple gravitational scattering due to
intervening clumps, the contribution of shear-rate to the distance
measure (the Weyl focusing) is comparable with one of matter inside the
beam (the Ricci focusing).
Taking into account the single gravitational lens effect, \cite{Wein76}
showed that an average flux from source in a clumpy universe with a low
deacceleration parameter $q_0$ is equal to the flux in the
Friedmann-Lema\^{i}re universe with the same $q_0$. \cite{Pea86}
developed Weinberg's argument to the case with an arbitrary $\Omega_0$
and the smoothness parameter $\bar{\alpha}\sim1$. Their results 
mean that the Dyer-Roeder distance multiplied by $\left<\mu\right>^{-1/2}$
($\left<\mu\right>$ is the averaged value of the gravitational
magnification factor $\mu$) can be considered to be equivalent to the
distance obtained by the Mattig formula. The averaged distance may be
useful in analyses such as $N-m$ or $N-z$ relations, in which the
gravitational lensing effect must be statistically taken into account
\citep[e.g.][]{OY90,YO92}. 
\par%
However, in the analysis of the distance-redshift relation for
individual objects the averaged distance is no longer available
\citep{HN86}. Then we are faced with the problem how  we should estimate
the distance to a specific object? We can assume that the distance may
be given by $\tilde{d}=\mu^{-1/2}d$ ($d$ denotes the Dyer-Roeder
distance) because the observed flux from the source is brighter by
factor $\mu$ than one with no lensing effects in the clumpy
universe. Although the above definition of the distance
$\tilde{d}=\mu^{-1/2}d$ seems reasonable, it is not a trivial problem to
show that the distance $\tilde{d}$ to the source is coincident with one 
obtained from the optical scalar equations, which is the subject that we 
will discuss in this paper. We will show that a light-path dependent
distance $\mu^{-1/2}d$ satisfies the Raychaudhuri equation for a null
geodesic and that the rates of expansion and shear in the scalar optics
theory can be expressed in terms of quantities of the gravitational lens
theory under the thin lens approximation. 
\par%
In this paper we assume that the universe is described, on average,  by
the Friedmann-Lema\^{i}tre model, but locally by the clumpy universe
given by \cite{DR73}. The same relation between the cosmological
time and the redshift\footnote{Usually the redshift is given by
$k^\mu u_\mu$, where $k^\mu$ and $u^\mu$ are the tangent to a null
geodesic and an observer's four velocity. Therefore the redshift of a
object depends on the distribution of inhomogeneities. However, here, we
ignore such dependence of the redshift \citep{DR74}.} is used in both
universe models (the homogeneous universe and the clumpy universe).
On the other hand, the distances to an object from an observer and
from another object  when gravitational lensing effects are
absent are assumed to be given, not by the Mattig formula, but by the
Dyer-Roeder distance. 
\par%
The organization of this paper is as follows. After a brief review of
the multiple gravitational lens effect in \S\S2.1, we give a formulation
of the continuous limit of the multiple gravitational lens effect
(\S\S2.2). In \S\S3.1, a light-path dependent angular diameter distance
$\tilde{d}$ will be defined, and we will obtain equations satisfied by
the elements of the Jacobian matrix $\mathbfss{A}$ which gives a mapping
from an observer plane to a source plane at redshift $z$. 
 In \S\S3.2, we will investigate  relations between
quantities in the gravitational lens theory and those in the scalar
optics theory. Finally, we shall give the summary and conclusion of this
paper. 
\section{Basic Formulation}
In this section we will present a brief review of multiple 
lensing effects and give a formulation of the continuous limit of the 
multiple lensing. Furthermore we will derive an equation satisfied by
the Jacobian matrix of the multiple gravitational lensings. 
\par
We assume that our universe is described by the Friedmann-Lema\^{i}re 
model in the very large scale (see Appendix A1), but  by the
clumpy model \citep{DR72,DR73} in the small scale. The fraction of the
smoothly distributed matter to the average density $\bar{\rho}$ of the
universe is $\bar{\alpha}$  (the smoothness parameter), and the rest
part of the matter in the universe is assumed to be concentrated into
the clumps which act as the gravitational lenses. In this model the
angular diameter distance $D(z_1;z_2)$ from a specific object at
redshift $z_1$ to another at $z_2$ is given by the Dyer-Roeder angular
diameter distance $D(z_{1};z_{2})$ (see Appendix A2). We introduce an
dimensionless angular diameter distance from $z_1$ to $z_2$ as
$d(z_1;z_2)=D(z_1;z_2)/(c/H_0)$, where $H_0$ is the Hubble constant. 
\subsection{Multiple lens equations}
Suppose that there are $N$ lenses which are randomly distributed at
redshift $z_i(0\le z_1<z_2<\cdots<z_N)$ in the universe and that  
$z_\mathrm{S}=z_{N+1}>z_N$ is a redshift of a source. We set lens planes  
perpendicular to the line of sight at each redshift. The origin on each 
lens plane is located on the point intersected with the line of sight.
\par
The multi-plane lens equation for the source is given by 
\citep[see, e.g.][]{SEF92,YNO04} 
\begin{equation}
\bmath{\theta}_S=\bmath{\theta}_1-\sum_{j=1}^N{d(z_j;z_\mathrm{S})\over 
d(0;z_\mathrm{S})}\bmath{\alpha}_j(D(0;z_j)\bmath{\theta}_j) 
\label{eq:mlpS}
\end{equation}
where $\bmath{\theta}_j$ is the angular position on the $i$-th lens
plane of the light from the source, and 
$\bmath{\alpha}_j$ denotes 
the deflection angle due to the $i$-th lens. It is useful to introduce a
renormalized deflection angle $\tilde{\bmath{\alpha}}_j$ defined by
$\tilde{\bmath{\alpha}}_j=d(0;z_j)\bmath{\alpha}_j$. In equation
(\ref{eq:mlpS}) $\bmath{\theta}_i$ are recursively expressed as follows: 
\begin{eqnarray}
\bmath{\theta}_i=\bmath{\theta}_1
 -\sum_{j=1}^i{d(z_j;z_i)\over d(0;z_j)d(0;z_i)}
\tilde{\bmath{\alpha}}_j(\bmath{\theta}_j),&&\quad\mbox{for $N\ge i>2$}
\label{eq:itheta}\\
\bmath{\theta}_2=\bmath{\theta}_1-{d(z_2;z_1)\over d(0;z_1)d(0;z_2)}
\tilde{\bmath{\alpha}}_1(\bmath{\theta}_1).
\hspace{3.5mm}&&\quad\mbox{for $i=2$}
\end{eqnarray}
From the above equations we see that $\bmath{\theta}_i$ in equation
(\ref{eq:itheta}) can be regarded as a source at redshift $z_i$ for the
foreground lenses. Furthermore, we can reduce a relation between
$\bmath{\theta}_{i-1},  \bmath{\theta}_i$ and $\bmath{\theta}_{i+1}$ as
follows: 
\begin{equation}
{\bmath{\theta}_{i+1}-\bmath{\theta}_i\over\chi_i-\chi_{i+1}}
-{\bmath{\theta}_i-\bmath{\theta}_{i-1}\over\chi_{i-1}-\chi_i}
=-(1+z_i)\tilde{\bmath{\alpha}}_i(\bmath{\theta}_i),
\label{eq:mlp}
\end{equation}
where $\chi_i$ is defined in the Appendix A3. The same equation is also
derived from the time delay of the multiple lensing effects with the
Fermat principle \citep{BL86}.
\par
The deflection angle $\tilde{\bmath{\alpha}}_i$ due to the $i$-th lens
in the clumpy universe is given by
\begin{equation}
\tilde{\bmath{\alpha}}_i(\bmath{\theta}_i)=\frac{3}{2}\Omega_0
{(1+z_i)^2d^2(0;z_i)\Delta z_i\over Y(z_i)}
{1\over\pi}\iint_{\cal D}d^2\bmath{\theta}' 
\Delta_{\bar{\alpha}}(\bmath{\theta}',Z(z_i))
{\bmath{\theta}_i-\bmath{\theta}'\over|\bmath{\theta}_i-\bmath{\theta}'|^2},
\label{eq:a_defl}
\end{equation}
where $\Delta z_i=z_i-z_{i-1}$, $Z(z)=c[T_0-T(z)]$ ($T_0$ and $T(z)$ are
the present cosmological time and a cosmological time when the light
emitted from the source at $z$, respectively) and 
$\Delta_{\bar{\alpha}}(\bmath{\theta}',Z(z_i))$ defined by
\begin{equation}
\delta_{\bar{\alpha}}\rho(D(0;z)\bmath{\theta},Z(z))\equiv
\rho(D(0;z)\bmath{\theta},Z(z))-\bar{\alpha}\bar{\rho}(z)
\equiv\rho_0(1+z)^3\Delta_{\bar{\alpha}}(\bmath{\theta},z)
\label{eq:deltarho}
\end{equation}
is an inhomogeneity of the matter distribution on the $i$-th lens plane.
\par
From equation (\ref{eq:mlp}) it can be shown that the Jacobian matrix 
$\mathbfss{A}_i={\partial\bmath{\theta}_i}/\partial\bmath{\theta}_1$ 
satisfies the differential equation
\begin{equation}
{\mathbfss{A}_{i+1}-\mathbfss{A}_i\over\chi_i-\chi_{i+1}}
-{\mathbfss{A}_i-\mathbfss{A}_{i-1}\over\chi_{i-1}-\chi_i}
=-(1+z_i)\tilde{\mathbfss{U}}_i(\bmath{\theta}_i)\mathbfss{A}_i,
\label{eq:recMag}
\end{equation}
where $\tilde{\mathbfss{U}}_i$ is defined as
\begin{eqnarray}
\tilde{\mathbfss{U}}_i(\bmath{\theta}_i)&=&
{\partial \bmath{\alpha}_i(\bmath{\theta}_i)\over\partial\bmath{\theta}_i}
={3\over2}\Omega_0
{(1+z_i)^2d^2(0;z_i)\Delta z_i\over Y(z_i)}
{1\over\pi}\iint_{\cal D}d^2\bmath{\theta}'
\Delta_{\bar{\alpha}}(\bmath{\theta}',Z(z_i))
\tilde{\mathbfss{U}}'(\bmath{\theta}_i-\bmath{\theta}'),
\label{eq:Ui}\\
\tilde{\mathbfss{U}}'(\bmath{\eta})&\equiv&
\left(
 \begin{array}{cc}
  \pi\delta^{(2)}(\bmath{\eta})-\Gamma_x(\bmath{\eta})
   &-\Gamma_y(\bmath{\eta})\\
  -\Gamma_y(\bmath{\eta})
   &\pi\delta^{(2)}(\bmath{\eta})+\Gamma_x(\bmath{\eta})
 \end{array}
\right). \label{eq:Utilde}
\end{eqnarray}
In equation (\ref{eq:Utilde}) $\bmath{\Gamma}=(\Gamma_x,\Gamma_y)$
denotes the shear-term due to the $i$-th lens, and is given by
\begin{equation}
\Gamma_x(\bmath{\eta})={\eta_x^2-\eta_y^2\over|\bmath{\eta}|^4},\qquad 
\Gamma_y(\bmath{\eta})={2\eta_x\eta_y\over|\bmath{\eta}|^4}.
\label{eq:Gammaxy}
\end{equation}
\cite{SeitzS92}  derived a recursive formula of the Jacobian matrix
$\mathbfss{A}_i$ equivalent to equation (\ref{eq:recMag}). However, their
formula is not useful to derive a differential equation of the Jacobian
matrix by taking a continuous limit, as mentioned below.
\subsection{Continuous limit of the multiple lens effect}
In above equations (\ref{eq:mlp}) and (\ref{eq:recMag}), we take the limit 
of $z_{i+1}\to z_i (\Delta z_i\to0)$ and obtain the following 
differential equations for $\bmath{\theta}(z)$ and $\mathbfss{A}(z)$ 
\begin{eqnarray}
{d\over dz}\left[{dz\over d\chi}{d\over dz}\bmath{\theta}(z)\right]
&=&\frac{3}{2}\Omega_0
{(1+z)^3d^2(0;z)\over Y(z)}{1\over\pi}\iint_{\cal D}d^2\bmath{\theta}'
\Delta_{\bar{\alpha}}(\bmath{\theta}',Z(z))
{\bmath{\theta}(z)-\bmath{\theta}'\over|\bmath{\theta}(z)-\bmath{\theta}'|^2},
\label{eq:cmlp} \\
{d\over dz}\left[{dz\over d\chi}{d\over dz}
\mathbfss{A}(\bmath{\theta}(z))\right]&&\nonumber\\
&&\hspace{-30mm}
=
\left[
 \frac{3}{2}\Omega_0
 {(1+z)^3d^2(0;z)\over Y(z)}{1\over\pi}\iint_{\cal D}d^2\bmath{\theta}'
 \Delta_{\bar{\alpha}}(\bmath{\theta}',Z(z))
 \tilde{\mathbfss{U}}'(\bmath{\theta}(z)-\bmath{\theta}')
\right]\mathbfss{A}(\bmath{\theta}(z)).
\label{eq:cMag}
\end{eqnarray}
\par
The equation (\ref{eq:cmlp}) has a formal solution
\begin{equation}
\bmath{\theta}(z)=\bmath{\theta}_0-\frac{3}{2}\Omega_0
\int_0^zd\zeta
{(1+\zeta)^2d(0;\zeta)d(\zeta;z)\over d(0;z)Y(\zeta)}
{1\over\pi}\iint_{\cal D}d^2\bmath{\theta}'
\Delta_{\bar{\alpha}}(\bmath{\theta}',Z(\zeta))
{\bmath{\theta}(\zeta)-\bmath{\theta}'
      \over|\bmath{\theta}(\zeta)-\bmath{\theta}'|^2},
\label{eq:formaltheta}
\end{equation}
provided that $\bmath{\theta}(z)$ satisfies  initial conditions
$\bmath{\theta}(0)=\bmath{\theta}_0$ and
$d\bmath{\theta}/dz|_{z=0}=\bmath{0}$ (see Appendix B1). If the
intervening lenses distribution is known, $\bmath{\theta}(z)$ can be
obtained in the form of a power series with $\Omega_{0}$. A source at
$\bmath{\theta}(z)$ in the no lenses case
($\Delta_{\bar{\alpha}}(\bmath{\theta}',Z(z))=0$) would be detected in
the direction to the angular position $\bmath{\theta}_0$.
\par 
We can also obtain the following formal solution of equation 
(\ref{eq:cMag}) with the initial conditions
$\mathbfss{A}(\bmath{\theta}(0))=\mathbfss{I}$ ~($2 \times 2$  unit
matrix) and $d\mathbfss{A}/dz|_{z=0}=\mathbfss{O}$~(the $2 \times 2$
zero matrix), 
\begin{equation}
\mathbfss{A}(\bmath{\theta}(z))
=\mathbfss{I}-\frac{3}{2}\Omega_0
 \int_0^zd\zeta {d(0;\zeta)d(\zeta;z)(1+\zeta)^2\over d(0;z) Y(\zeta)}
 \hat{\mathbfss{U}}(\bmath{\theta}(\zeta),\zeta)\mathbfss{A}(\zeta),
\label{eq:sMag}
\end{equation}
where the matrix $\hat{\mathbfss{U}}$ is given in terms of its elements
as follows :  
\begin{eqnarray}
\hat{\mathbfss{U}}(\bmath{\theta}(z),z)&\equiv&
\frac{1}{\pi} \iint_{\cal D}d^2\bmath{\theta}'
\Delta_{\bar{\alpha}}(\bmath{\theta}',z)
\tilde{\mathbfss{U}}'(\bmath{\theta}(z)-\bmath{\theta}')
\nonumber\\ 
&\equiv& 
\left(
 \begin{array}{cc}
  \Delta_{\bar{\alpha}}(\bmath{\theta}(z),z)-\gamma_x(\bmath{\theta}(z),z)&
   -\gamma_y(\bmath{\theta}(z),z)\\
  -\gamma_y(\bmath{\theta}(z),z)&
   \Delta_{\bar{\alpha}}(\bmath{\theta}(z),z)+\gamma_x(\bmath{\theta}(z),z)\\
 \end{array}
	\right).
\label{eq:deltagamma}
\end{eqnarray}
In equation (\ref{eq:deltagamma}) $\bmath{\gamma}(\bmath{\theta}(z),z)$
is defined by 
\begin{equation}
\bmath{\gamma}(\bmath{\theta}(z),z)=\frac{1}{\pi}
\iint_{\cal D}d^2\bmath{\theta}'
\Delta_{\bar{\alpha}}(\bmath{\theta}',z)\bmath{\Gamma}
\left(\bmath{\theta}(z)-\bmath{\theta}'\right).
\label{eq:sgamma}
\end{equation}
As described in Appendix B2, equation (\ref{eq:sMag}) can be rewritten
in the form: 
\begin{equation}
\mathbfss{A}(\bmath{\theta}(z))=\mathbfss{I}-\frac{3}{2}\Omega_0
\int_{\chi(0)}^{\chi(z)}d\chi_1\int_{\chi(0)}^{\chi_1}d\chi_2
(1+\zeta_2)^5d^4(0;\zeta_2)
\hat{\mathbfss{U}}(\bmath{\theta}(\zeta_2),\zeta_2)
\mathbfss{A}(\bmath{\theta}(\zeta_2)).
\label{eq:SolMag}
\end{equation}
Using this expression, we can easily obtain the second derivative of 
$\mathbfss{A}$ with respect to $\chi(z)$: 
\begin{equation}
\mathbfss{A}''(\chi)=-\mathbfss{W}''(\chi)\mathbfss{A}(\chi),\label{eq:AWA}
\end{equation}
where the  matrix $\mathbfss{W}$ is given by
\begin{eqnarray}
\mathbfss{W}(\chi)
&\equiv&\frac{3}{2}\Omega_0\int_{\chi(0)}^{\chi(z)}d\chi_1
\int_{\chi(0)}^{\chi_1}d\chi_2(1+\zeta_2)^5d^4(0;\zeta_2)
\hat{\mathbfss{U}}(\bmath{\theta}(\zeta_2))\nonumber\\
&\equiv&
\left(\begin{array}{cc}
\kappa(\chi)-S_x(\chi) &-S_y(\chi)\\
       -S_y(\chi)& \kappa(\chi)+S_x(\chi)
\end{array}\right).
\label{eq:W}
\end{eqnarray}
In equation (\ref{eq:W}) the convergence $\kappa(\chi)$ and the shear
$\bmath{S}(\chi)$ produced by a single lensing are defined as follows: 
\begin{eqnarray}
\kappa(\chi)&=&\frac{3}{2}\Omega_0\int_{\chi(0)}^{\chi(z)}d\chi_1
\int_{\chi(0)}^{\chi_1}d\chi_2(1+\zeta_2)^5d^4(0;\zeta_2)
\Delta_{\bar{\alpha}}(\bmath{\theta}(\zeta_2)),\label{eq:cumuCon}\\
\bmath{S}(\chi)&=&\frac{3}{2}\Omega_0\int_{\chi(0)}^{\chi(z)}d\chi_1
\int_{\chi(0)}^{\chi_1}d\chi_2(1+\zeta_2)^5d^4(0;\zeta_2)
\bmath{\gamma}(\bmath{\theta}(\zeta_2)).\label{eq:cumuShear}
\end{eqnarray}
In equation (\ref{eq:AWA}), $\mathbfss{A}(\chi), \mathbfss{W}(\chi)$ and
the prime $~'~$ denote 
$\mathbfss{A}(\bmath{\theta}(z(\chi))),\mathbfss{W}(\bmath{\theta}(z(\chi)))$ 
and the derivative with respect to $\chi$, respectively. 
Appendix B2  gives a proof that equation
(\ref{eq:AWA}) is the equivalent to equation (\ref{eq:cMag}).  
In evaluation of $\mathbfss{A}$, we must notice that $\mathbfss{A}$
depends on $\bmath{\theta}(z)$ which is the solution of equation
(\ref{eq:cmlp}). 

\section{Magnification Factor and the Optical Scalar Equations}
\subsection{Jacobian matrix}
In the clumpy universe \citep{DR73} if the light ray is not affected by 
any clump which intervenes between a source and the observer, 
the flux $f_0$ from the source at redshift $z$ with absolute luminosity
$L$ would be detected as 
\begin{equation}
f_0={L\over4\pi(c/H_0)^2(1+z)^4d^2(0;z)}.
\end{equation}
However, in the real case, observers detect a light ray which is
gravitationally lensed by the clumps. Therefore the observed flux
$f_{\mathrm{obs}}$ is given by $f_\mathrm{obs}=\mu f_0$, where $\mu$ is
the gravitational magnification factor defined by
$\mu=\det\mathbfss{A}^{-1}$. It is important to notice that $\mu$ 
depends on the light-path. 
Since observers recognize that the angular diameter distance to the
source with flux $f_\mathrm{obs}$ is given by 
$(H_{0}/c)\times L^{1/2}[4\pi(1+z)^4f_\mathrm{obs}]^{-1/2}$, 
the new angular diameter distance $\tilde{d}$ is defined by
\begin{equation}
\tilde{d}(\bmath{\theta}(z)|0;z)\equiv \mu^{-1/2}(\bmath{\theta}(z))d(0;z).
\label{eq:newdis}
\end{equation}
We should notice that $\tilde{d}$ has a dependence on the
light-path. Then, even if two sources have a same redshift and a same 
absolute luminosity, their distances are not always reduced to be same
values. In the below we investigate the equation of the new distance
$\tilde{d}$. Hereafter we assume $\mu\ne0$, i.e. the light
ray considered here does not pass through a conjugate point
\citep{Wald84}. 
\par
For this purpose we need, first, to obtain an equation of the
gravitational magnification factor $\mu$. 
In general, Jacobian matrix $\mathbfss{A}$ from the observe plane to
a source plane at redshift $z$ is expressed as 
\begin{equation}
\mathbfss{A}=
\left( \begin{array}{cc}
K_x+G_x&G_y+K_y\\G_y-K_y&K_x-G_x\end{array}\right)
=K_x\sigma_0+G_y\sigma_1+iK_y\sigma_2+G_x\sigma_3,
\label{eq:generalA}
\end{equation}
where $\sigma_i$'s are the Pauli matrices:
\begin{equation}
  {\sigma}_0=\mathbfss{I},\
  {\sigma}_1=\left(\begin{array}{cc}0&1\\1&0\end{array}\right),\
  {\sigma}_2=\left(\begin{array}{cc}0&-i\\i&0\end{array}\right),\
  {\sigma}_3=\left(\begin{array}{cc}1&0\\0&-1\end{array}\right).
\end{equation}
In this expression, the magnification factor $\mu$ is given by
$\mu=\det\mathbfss{A}^{-1}=
(\bmath{K}\cdot\bmath{K}-\bmath{G}\cdot\bmath{G})^{-1}$, where 
$K_x, K_y$ and $\bmath{G}$ denote the cumulative convergence-, twist-
and shear-terms, respectively. In a single lensing case, $\mathbfss{A}$
is symmetric, i.e., the twist-term $K_y$ always vanishes. On the other
hand, in the multiple lensing case, it does not vanish, in general.
\par
According to \cite{SSE94}, we can define an optical deformation matrix
$\mathbfss{Q}$ in terms of $\mathbfss{A}$ and its derivative with
respect to $\chi$ as follows:
\begin{equation}
\mathbfss{Q}\equiv\mathbfss{A}'\mathbfss{A}^{-1},\label{eq:defQ}
\end{equation}
which gives 
\begin{equation}
\mathbfss{A}'=\mathbfss{Q}\mathbfss{A}.\label{eq:AQA}
\end{equation}
Inserting equation (\ref{eq:generalA}) into equation (\ref{eq:defQ}),
we obtain an explicit form of the optical deformation matrix:
\begin{eqnarray}
\mathbfss{Q}
&=&\left(\ln\mu^{-1/2}\right)'\sigma_0
+\varepsilon_{ab}\left(\hat{K}_a^{}\hat{G}_b'
	       -\hat{K}_a'\hat{G}_b^{}\right)\sigma_1
\nonumber\\
&&\hspace{2cm}
+i\varepsilon_{ab}\left(\hat{K}_a^{}\hat{K}_b'
	       +\hat{G}_a'\hat{G}_b^{}\right)\sigma_2
+\delta_{ab}\left(\hat{K}_a^{}\hat{G}_b'
	       -\hat{K}_a'\hat{G}_b^{}\right)\sigma_3,\label{eq:Qc}
\end{eqnarray}
where $\varepsilon_{xx}=\varepsilon_{yy}=0,\
\varepsilon_{xy}=-\varepsilon_{yx}=1$,
$\hat{K}_a\equiv\mu^{1/2}K_a$ and $\hat{G}_a\equiv\mu^{1/2}G_a$.
\par
Using a relation (\ref{eq:AWA}), the definition of $\mathbfss{Q}$ and
its derivative with respect to $\chi$, we can obtain a relation between
$\mathbfss{Q}$ and $\mathbfss{W}''$ as follows
\begin{equation}
\mathbfss{Q}'+\mathbfss{Q}^2=-\mathbfss{W}'',\label{eq:QW}
\end{equation}
 where $\mathbfss{W}''$ is written in terms of the Pauli matrices as
$\mathbfss{W}''=\kappa''\sigma_0-S_y''\sigma_1-S_x''\sigma_3$.
Using equations (\ref{eq:Qc}) and (\ref{eq:QW}), we find equations of the
coefficients of the Pauli matrices in $\mathbfss{Q}$ in the following
\begin{eqnarray}
&&\left(\ln\mu^{-1/2}\right)''+\left[\left(\ln\mu^{-1/2}\right)'\right]^2
+\left\{
\left[\varepsilon_{ab}\left(\hat{K}_a^{}\hat{G}_b'
	       -\hat{K}_a'\hat{G}_b^{}\right)\right]^2
\right.
\nonumber\\
&&
\left.
+\left[\delta_{ab}\left(\hat{K}_a^{}\hat{G}_b'
	       -\hat{K}_a'\hat{G}_b^{}\right)\right]^2
-\left[\varepsilon_{ab}\left(\hat{K}_a^{}\hat{K}_b'
	       +\hat{G}_a'\hat{G}_b^{}\right)\right]^2
\right\}=-\kappa'',\label{eq:eqmu}\\
&&
\left[\delta_{ab}\left(\hat{K}_a^{}\hat{G}_b'
	       -\hat{K}_a'\hat{G}_b^{}\right)\right]'
+2\left(\ln\mu^{-1/2}\right)'\left[\delta_{ab}\left(\hat{K}_a^{}\hat{G}_b'
	       -\hat{K}_a'\hat{G}_b^{}\right)\right]=S_x'',\label{eq:eqSx}\\
&&
\left[\varepsilon_{ab}\left(\hat{K}_a^{}\hat{G}_b'
	       -\hat{K}_a'\hat{G}_b^{}\right)\right]'
+2\left(\ln\mu^{-1/2}\right)'\left[\varepsilon_{ab}\left(\hat{K}_a^{}\hat{G}_b'
	       -\hat{K}_a'\hat{G}_b^{}\right)\right]=S_y'',\label{eq:eqSy}\\
&&
\left[\varepsilon_{ab}\left(\hat{K}_a^{}\hat{K}_b'
	       +\hat{G}_a^{}\hat{G}_b'\right)\right]'
+2\left(\ln\mu^{-1/2}\right)'\left[\varepsilon_{ab}\left(\hat{K}_a^{}\hat{K}_b'
	       +\hat{G}_a^{}\hat{G}_b'\right)\right]=0.\label{eq:eqrotate}
\end{eqnarray}
Equation (\ref{eq:eqmu}) is equivalent to one derived by \cite{SSE94}.
They gave the equation of the magnification factor in the clumpy
universe by using a rate of shear derived from a Weyl term of the scalar
optics theory in the linearized general relativity. On the other hand,
equation (\ref{eq:eqmu}) is derived  by taking the continuous limit of
the multiple gravitational lens theory. It should be noticed, therefore,
that in our formulation we can obtain the right hand side of equation
(\ref{eq:eqmu}) from data of lens distributions itself, without
referring to the Weyl term (the metric terms given by solving the
Einstein equation).   
\par
The second derivative of $\tilde{d}$ with respect to the affine
parameter $v$ can be expressed as combination of the second derivative of
$d$ with respect to $v$ and the second derivative of
$\mu^{-1/2}$ with respect to $\chi$:
\begin{equation}
{d^2\ \over dv^2}\tilde{d}(\bmath{\theta}(z)|0;z)
=\mu^{-1/2}{d^2\ \over dv^2}d(0;z)
  +{1\over d^3(0;z)}{d^2\ \over d\chi^2}\mu^{-1/2}(\bmath{\theta}(z)).
\label{eq:newdchi}
\end{equation}
Combining equation (\ref{eq:eqmu}) with equation (\ref{eq:newdchi}), the
Dyer-Roeder equation (\ref{eq:DREQ}) yields  
\begin{equation}
{d\over dv}\tilde{\Theta}+\tilde{\Theta}^2+
\left(\delta_{ab}\tilde{\Sigma}_{ab}\right)^2
+\left(\varepsilon_{ab}\tilde{\Sigma}_{ab}\right)^2
-\left(\varepsilon_{ab}\tilde{\omega}_{ab}\right)^2
=-{4\pi G\over H_0}\rho(D(0;z)\bmath{\theta}(z),Z(z))(1+z)^2,
\label{eq:eqTheta}
\end{equation}
where
\begin{eqnarray}
\tilde{\Theta}&=&{d\over dv}\ln\tilde{d}(\bmath{\theta}(z)|0;z),
\label{eq:Theta}\\
\tilde{\Sigma}_{ab}&=&
 -{\hat{K}_a^{}\hat{G}_b'-\hat{K}_a'\hat{G}_b\over d^2(0;z)},
\label{eq:Sig}\\
\tilde{\omega}_{ab}&=&
 -{\hat{K}_a^{}\hat{K}_b'+\hat{G}_a^{}\hat{G}_b'\over d^2(0;z)}.
\label{eq:ome}
\end{eqnarray}
Furthermore, inserting equations (\ref{eq:Sig}), (\ref{eq:ome}) into
equations (\ref{eq:eqSx})-(\ref{eq:eqrotate}), we find that
\begin{eqnarray}
&&
{d\ \over dv}\left(
\delta_{ab}\tilde{\Sigma}_{ab}\right)+2\tilde{\Theta}\delta_{ab}
\tilde{\Sigma}_{ab}
={4\pi G\over H_0^2}\bar{\rho}(z)(1+z)^2\gamma_{x}(\bmath{\theta}(z)),
\label{eq:eqSigmax}\\
&&{d\ \over dv}\left(\varepsilon_{ab}\tilde{\Sigma}_{ab}\right)
+2\tilde{\Theta}\varepsilon_{ab}\tilde{\Sigma}_{ab}
={4\pi G\over H_0^2}\bar{\rho}(z)(1+z)^2\gamma_{y}(\bmath{\theta}(z)),
\label{eq:eqSigmay}\\
\noalign{\hbox{and}}
&&{d\ \over dv}
\left(\varepsilon_{ab}\tilde{\omega}_{ab}\right)
+2\tilde{\Theta}\varepsilon_{ab}\tilde{\omega}=0.
\label{eq:eqomega}
\end{eqnarray}
Here we define
$\tilde{\Sigma}=(\delta_{ab}+i\varepsilon_{ab})\tilde{\Sigma}_{ab}$ 
and $\tilde{\omega}=\varepsilon_{ab}\tilde{\omega}_{ab}$ and then
rewrite equations (\ref{eq:eqTheta}),
(\ref{eq:eqSigmax})-(\ref{eq:eqomega}) as follows:
\begin{eqnarray}
&&{d\ \over dv}\tilde{\Theta}+\tilde{\Theta}^2
+|\tilde{\Sigma}|^2-\tilde{\omega}^2=\tilde{\cal R},\label{eq:expan}\\
&&{d\ \over dv}\tilde{\Sigma}+2\tilde{\Theta}\tilde{\Sigma}
=\tilde{\cal F},\label{eq:shear}\\
&&{d\ \over dv}\tilde{\omega}+2\tilde{\Theta}\tilde{\omega}=0,
\label{eq:rotation}
\end{eqnarray}
where 
\begin{eqnarray}
\tilde{\cal R}&=&-{4\pi G\over H_0^2}\rho(D(0;z)\bmath{\theta}(z)|0;z)(1+z)^2,
\label{eq:Ricci}\\
\tilde{\cal F}&=&{4\pi G\over H_0^2}
\bar{\rho}(z)\left[\gamma_x(\bmath{\theta}(z))+i\gamma_y(\bmath{\theta}(z))
\right](1+z)^2\nonumber\\
&=&{4G\over H_0^2}\iint d^2\bmath{\theta}'
\delta_{\bar{\alpha}}\rho(D(0;z)\bmath{\theta}',Z(z))
\Gamma(\bmath{\theta}(z)-\bmath{\theta}')(1+z)^2,
\label{eq:Weyl}
\end{eqnarray}
and $\Gamma(z)=\Gamma_x(z)+i\Gamma_y(z)$ denotes the complex shear term
due to a 
lens at redshift $z$. 
\par
From  equations (\ref{eq:Theta}) and (\ref{eq:expan}), we find that our
new distance (\ref{eq:newdis}) satisfies the following equation:
\begin{equation}
{d^2 \over dv^2}\tilde{d}(\bmath{\theta}(z)|0;z)
+\left[
  \left|\tilde{\Sigma}\right|^2
  +{4\pi G\over H_0^2}\rho(D(0;z)\bmath{\theta}(z),Z(z))(1+z)^2
\right]\tilde{d}(\bmath{\theta}(z)|0;z)=0.\label{eq:Eqdtilde} 
\end{equation}
We should notice that we cannot obtain solution of equation
(\ref{eq:Eqdtilde}) without solving equation (\ref{eq:cmlp}). 

\subsection{Relations between the Jacobian matrix and the optical scalars}
Since the cross-sectional area of the light bundle, ${\cal A}$, is
proportional to square of the angular diameter distance \citep{DR72},
we can rewrite $\tilde{\Theta}$ as
\begin{equation}
\tilde{\Theta}=\frac{1}{2}{d\over dv}\ln{\cal A}, \label{eq:calAtoTheta}
\end{equation}
which is the rate of expansion in the scalar optics theory. Furthermore,
equation (\ref{eq:Ricci}) can be regarded as the Ricci term, which is
given by $\tilde{\cal R}=-(1/2)R_{\mu\nu}k^{\mu}k^{\nu}$. Therefore, we
can show that  equation (\ref{eq:eqTheta}) or (\ref{eq:expan}) can be
regarded as the 
optical scalar equation of the rate of expansion, if $\tilde{\Sigma}$ and 
$\tilde{\omega}$ correspond to the rates of shear and rotation in the
scalar optics theory. 
\par
First, let us investigate $\tilde{\omega}$. Combining initial conditions
of $\mathbfss{A}$ and definitions of $\tilde{\Theta}$ and
$\tilde{\omega}$, we obtain a solution of (\ref{eq:rotation}),
$\tilde{\omega}=0$. In the scalar optics, the rotation rate
$\omega\equiv\sqrt{\omega_{\alpha\beta}\omega^{\alpha\beta}/2}$
always vanishes, because the covariant derivative of the
tangent vector of the null geodesic is symmetric, i.e.,
$\omega_{\alpha\beta}=\nabla_{[\alpha}{}_,k_{\beta]}=0$. From this fact
and equation (\ref{eq:rotation}), we can regard $\tilde{\omega}$ as the
rotation rate in the scalar optics theory. 
Furthermore, we find that the optical deformation matrix $\mathbfss{Q}$
is symmetric. In a discrete multiple gravitational lensing theory, the
symmetry of a matrix corresponding to the optical deformation matrix
was shown by \cite{SeitzS92} by means of their recursive formula of the
Jacobian matrix. However, in their paper, the reasons of the symmetry
was not clarified.
\par
Next, let us notice the right hand side of equation (\ref{eq:shear}),
$\tilde{\cal F}$. This is similar to the Weyl term in the scalar optics theory. 
The difference between $\tilde{\cal F}$ and the Weyl term in the linear
perturbation theory of the general relativity \citep{Sasaki87, FS89}
exists in the following point: 
while the Weyl term at arbitrary
point $(D(0;z)\bmath{\theta}(z),Z(z))$ on the light-path in the linear
perturbation theory  has 
some contribution from the matter  
in the whole universe  outside the light bundle, 
$\tilde{\cal F}$ in our formalism has a contribution from only the matter on
the lens plane with the same redshift $z$. This  comes from
the fact that our formalism is based on the thin lens
approximation. Therefore, we can regard $\tilde{\cal{F}}$ as the Weyl term in
scalar optics theory as long as the thin lens approximation is valid.  
\par
Finally we find that equations (\ref{eq:expan})--(\ref{eq:rotation}) are
equivalent to the optical scalar equations (\ref{eq:opS}) in an 
inhomogeneous universe under the thin lens approximation.
Consequently the optical scalars are expressed  in terms of the elements
of the Jacobian matrix and the Dyer--Roeder angular diameter distance as
follows:
\begin{equation}
\Theta=
{d\ \over dv}\ln\left[\mu^{-1/2}d(0;z)\right],\
\Sigma=-\left(\delta_{ab}+i\epsilon_{ab}\right)
\left[{\hat{K}_a\hat{G}_b'-\hat{K}_a'\hat{G}_a\over d^2(0;z)}\right],\ 
\omega=-\epsilon_{ab}
\left[{\hat{K}_a\hat{K}_b'+\hat{G}_a'\hat{G}_a\over d^2(0;z)}\right],
\label{eq:LopS}
\end{equation}
where the last equation is exactly zero.
\par
\section{Summary and Conclusions}
In this paper we have formulated the continuous limit of the multiple
gravitational lens effect by taking the limit in which the number of
lens planes is infinite and then the redshift-intervals  of lenses are
infinitesimal with keeping the present average density $\bar{\rho}_0$
fixed. It, however, does not mean that the number of lenses must be 
infinite.  In this case, 
we can interpret that the finite number of lenses have some finite 
masses in their lens planes and other infinite number of lenses have 
zero masses in the planes, and that in the zero-mass lens planes, the 
deflection angles, $\tilde{\bm{\alpha}}$, should be $\bm{0}$ and the 
matrices, $\tilde{\mathbfss{U}}$, should be $\mathbfss{I}$. The
formalism is able to be adopted in the case where there  are some finite
number of lenses between sources and us. 
\par
Using the formalism 
we found that the new angular diameter distance, 
$\tilde{d}=\mu^{-1/2}d$ satisfies the optical 
scalar equation of the expansion rate $\tilde{\Theta}$ in an
inhomogeneous universe. It is interesting that the influence of 
inhomogeneities on the angular diameter distance is completely recovered 
by multiplication of $\mu^{-1/2}$ to the Dyer--Roeder distance, which is 
obtained by ignoring the influence of all clumps on the light beams. 
\par
We also found that quantities defined in the
scalar optics theory (the rates of expansion $\Theta$, shear $\Sigma$ and
rotation $\omega$) are expressed in terms of quantities defined in the
gravitational lens theory (the terms of convergence $K_x$, twist $K_y$ and
shear $\bmath{G}$). 
\cite{SeitzS94} showed that the absolute square of the shear rate,
$\left|\Sigma\right|^2$, can be expressed by some combination of the
determinants and traces of the Jacobian matrices in the discrete
multiple gravitational lensing theory. It is equivalent to
one in our formalism, which is also written as
\[
 \left|\tilde{\Sigma}\right|^2={\hat{\bmath{G}}'\cdot\hat{\bmath{G}}'
-\hat{\bmath{K}}'\cdot\hat{\bmath{K}}'\over d^4(0;z)}.
\]
This expression gives an immediate estimation of the shear rate by
tracing the lens equation (\ref{eq:cmlp}) from  the raw 
data of local distribution of matters.
\par
The formalism includes the framework of the weak gravitational lensing
theory \citep[e.g.][]{Bart01}, in which the smoothness parameter
$\bar{\alpha}$ is assumed to be unity(the distance measure is
given by the Mattig formula and only single gravitational lensing is
taken in account). In their framework,  when the
light ray propagates through some underdense region, 
$\rho <\bar{\rho}$, the light is assumed to be deflected by a lens with some
negative mass (deflecting mass $\delta\rho=\rho-\bar{\rho}<0$). 
Therefore such lenses with negative masses give some brightening effect
similar to the usual clumps on the flux when the light beam passes away
from such lenses, because the shear rate due to the negative mass lenses 
contributes to the Raychaudhuri equation in the form of
$|\delta\rho|^2$. Then there is some possibility that the weak lensing
theory does overestimate the distance measure due to the negative lenses
unless the convergence due to the negative mass lens in the light beam 
balances with the shear outside the light beam due to the other ones.
In our formalism , unlike the weak lensing theory, we do not need to
introduce such lenses with the negative masses in order to take into
account the underdense regions in which the matter density is only
considered to be $\rho=\bar{\alpha}\bar{\rho}$. Therefore there is not
the above possibility because the shear in such underdense region does
not contribute to the distance measure.
\par
Finally, we comment on equation (\ref{eq:Eqdtilde}) which is the
generalized Dyer-Roeder equation. \cite{SEF92} also mentioned a similar
equation, which is slightly different from ours. In their equation, the
density of clumps
$\delta_{\bar{\alpha}}\rho=\rho-\bar{\alpha}\bar{\rho}$ is not
considered, because they assumed that the light beam does not pass
through bound clumps. However, among real situations, there are cases in
which the contributions of both the matter density of the bound clump in
the light bundle and the shear rate of the light bundle are compatible
\citep{WS90}. Even in such intermediate case, equation (\ref{eq:Eqdtilde})
gives the angular diameter distance to a target source, by solving the
lens equations (\ref{eq:cmlp}) based on observed distribution of
foreground clumps. 
\section*{Acknowledgments}
This research was partially supported by the Grant-in-Aid for Scientific
Research on Priority Areas (13135218) of the Ministry of Education,
Science, Sports and Culture of Japan.
\appendix
\section{Relations between redshift 
$\bmath{\lowercase{z}}$, affine parameter $\bmath{\lowercase{v}}$ and 
\mbox{$\chi$}-function}
In this appendix, we briefly describe the relation between $z$ (redshift), 
$v$(affine parameter) and $\chi$ ($\chi$-function). 
\subsection{Background universe}
We assume that the universe is described by the Friedmann-Lema\^{\i}tre 
universe on average. Then the geometry on average is expressed as 
\begin{equation}
ds^2=c^2dT^2-a^2(T)\left[
{dr^2\over1-kr^2}+r^2\left(d\theta^2+\sin^2\theta d\phi^2\right)
\right],
\label{eq:FLgeom}
\end{equation}
where $T$ is the cosmological time, $a(T)$ denotes the scale factor of 
the universe which has the dimension of distance and $k$ is related with 
the curvature of the universe. In this geometry, the Einstein field 
equations can be written as
\begin{eqnarray}
\left({\dot{a}\over a}\right)^2&=&{8\pi G\over3}\bar{\rho}
+{\Lambda c^2\over 3}-{kc^2\over  a^2},\label{eq:Hubble}\\
\Lambda c^2&=&2{\ddot{a}\over a}+\left({\dot{a}\over a}\right)^2
+{kc^2\over a^2}, \label{eq:decc}
\end{eqnarray}
where $\bar{\rho}$ and $\Lambda$ are the mean density of the universe 
and the cosmological constant, respectively, and the dot denotes the 
derivative with $T$. Since $a(T)$ and $\bar{\rho}$ are given in terms of 
their present values $a_0$ and $\bar{\rho}_0$ by $a(T)=a_0/(1+z)$ and 
$\bar{\rho}=\bar{\rho}_0(1+z)^3$ (in the matter-dominated era), 
respectively, equations (\ref{eq:Hz}) and (\ref{eq:decc}) are written as 
\begin{eqnarray}
\left({\dot{a}\over a}\right)^2&=&H_0^2\left[
\Omega_0(1+z)^3+\lambda_0-K(1+z)^2
\right]\equiv H_0^2Y^2(z),\label{eq:Hz}\\
-\left({\ddot{a}a\over\dot{a}^2}\right)&=&q(z)
={\Omega_0(1+z)^3/2-\lambda_0\over Y^2(z)},\label{eq:qz}
\end{eqnarray}
where $H_0$ is the present Hubble constant,
$\Omega_0=8\pi G\bar{\rho}_0/3H_0^2$, 
$\lambda_0=\Lambda c^2/3H_0^2$ and $K=kc^2/a_0^2H_0^2$.
The above equations gives relations between $\Omega_0, \lambda_0$ and 
$K$:
\begin{eqnarray}
&K=\Omega_0+\lambda_0-1,&\label{eq:KOmLam}\\
&q_0=\frac{1}{2}\Omega_0-\lambda_0,&\label{eq:q0}
\end{eqnarray}
where $q_0$ is the present deacceleration parameter defined as 
$-(\ddot{a}a/\dot{a}^2)_0$. 
It follows from equation (\ref{eq:Hz}) that the relation between $T$ and 
$z$ is given by: 
\begin{equation}
-cdT={c\over H_0}{dz\over(1+z)Y(z)}.
\end{equation}
\par
Furthermore, in the Friedmann-Lema\^{\i}tre universe, an affine
parameter $v$ defined  
along a null geodesic is given in terms of redshift $z$ by
\begin{equation}
dv={dz\over(1+z)^2Y(z)}.\label{eq:vz}
\end{equation}
In this universe, the angular diameter distance $D_\mathrm{FL}(z_2;z_1)$ 
from $z_1$ to $z_2$ is given as
\begin{equation}
D_\mathrm{FL}(z_2;z_1)\equiv a(T_1)r(z_2;z_1)
={c\over H_0}{1\over\sqrt{K}(1+z_1)}
\sin\left[
\sqrt{K}\int_{z_2}^{z_1}{dz\over Y(z)}\right].\label{eq:disFL}
\end{equation}
This is the Mattig formula.
\subsection{The Dyer-Roeder distance}
\cite{Sachs61} gave geometrical equations of light bundle propagating in 
the general space-time from the geodesic deviation for the light ray as 
follows: 
\begin{equation}
{d\ \over dv}\Theta+\Theta^2+|\Sigma|^2={\cal R},\qquad 
{d\ \over dv}\Sigma+2\Sigma\Theta={\cal F},\label{eq:opS}
\end{equation}
where $\Theta$ and $\Sigma$ denote the rates of expansion and shear
factor of the light bundle, ${\cal R}$ and ${\cal F}$ are the Ricci and 
Weyl terms. The rate of expansion $\Theta$ is defined as 
$\Theta=d\ln\sqrt{\cal A}/dv$, where ${\cal A}$ denotes the 
cross-sectional area of the light bundle. Since the angular diameter 
distance $D$ to a source at redshift $z$ is proportional to 
$\sqrt{\cal A}$, equations (\ref{eq:opS}) are rewritten as
\begin{equation}
{d^2\ \over dv^2}D=\left({\cal R}-|\Sigma|^2\right)D, \qquad
{1\over D^2}{d\ \over dv}\left(D^2\Sigma\right)={\cal F}.
\label{eq:OSEQ}
\end{equation}
In general it follows that the angular diameter distance depends not 
only on the Ricci term but also on the Weyl term, where the Weyl term is 
due to the inhomogeneities outside  the light bundle while the Ricci 
term originates from the matter density in the light bundle,
\par
\cite{DR72,DR73} assumed that for the light beams far away 
from all the inhomogeneities, the contribution of $\Sigma$ to the 
distance $D$ from the source is negligible and that the matter-density 
in the light bundle is given by $\bar{\alpha}\bar{\rho}$, where 
$\bar{\alpha}$ (smoothness parameter) is the fraction of the density 
smoothly distributed in the universe to the mean density of the universe, 
and derived the angular diameter distance in the clumpy universe. which 
satisfies the following equation:
\begin{equation}
{d^2\ \over dv^2}D+{4\pi G\over H_0^2}\bar{\alpha}\bar{\rho}(1+z)^2D=0.
\label{eq:DREQ}
\end{equation}
The relation (\ref{eq:vz}) between the affine parameter along the light 
ray and redshift allows us to express equation (\ref{eq:DREQ}) as follows:
\begin{equation}
{d\over dz}\left[
	      (1+z)^2Y(z){d\ \over dz}
	    \right]d(z_0;z)
	    +{3\over2}\Omega_0\bar{\alpha}{(1+z)^3\over Y(z)}d(z_0;z)=0,
\end{equation}
where $d(z_0;z)$ denotes the dimensionless angular diameter distance 
$D(z_0;z)/(c/H_0)$ from an object at redshift $z_0$ to another at 
$z (z\le z_0)$.  The initial conditions at $z=z_0$ are given by
\begin{eqnarray}
&&\hspace{-6mm}d(z_0;z)\Bigr|_{z=z_0}=0,\nonumber\\
&&\hspace{-6mm}{d\ \over dz}d(z_0;z_b)\Bigr|_{z=z_0}={1\over(1+z_0)Y(z_0)},
\label{eq:Hubblelaw}
\end{eqnarray}
\citep{DR73}, where the last condition is the Hubble law at redshift
$z_0$. In the case with $\bar{\alpha}=1$, equation (\ref{eq:disFL}) is a
solution of the Dyer-Roeder equation. 
\par
\subsection{The \mbox{$\chi$}-function}
The $\chi$-function is introduced by \cite{SEF92} as follows
\begin{equation}
\chi(z)\equiv\int_z^\infty{dz\over(1+z)^2Y(z)d^2(0;z)},\label{eq:chifunc}
\end{equation}
where $d(0;z)$ is the dimensionless Dyer-Roeder angular diameter distance.
The relation between  $\chi$, $v$ and $z$ is then expressed as
\begin{equation}
d\chi=-{dv\over d^2(0;z)}=-{dz\over(1+z)^2Y(z)d^2(0;z)}.\label{eq:chivz}
\end{equation}
Accordingly we can use the $\chi$-function as a variable instead of
redshift $z$ or affine parameter $v$. 
\par
The deference between values of $\chi$ at redshifts $z_i$ and at $z_j$ 
($z_i<z_j$) has an interesting relation to the Dyer-Roeder angular 
diameter distance:
\begin{equation}
\chi(z_i;z_j)\equiv\chi(z_i)-\chi(z_j)=
{d(z_i;z_j)\over(1+z_i)d(0;z_i)d(0;z_j)}.
\label{eq:chiddd}
\end{equation}
In the rest part of this subsection we will prove this relation. First 
of all, we investigate the differential equation which
$d(0;z)\chi(z_i;z)$ satisfies. Let $f(z)$ denote $d(0;z)\chi(z_i;z)$ and 
differentiate it with respect to affine parameter $v$:
\begin{equation}
 {d\ \over dv}f(z)=\left[{d\ \over dv}d(0;z)\right]\chi(z_i;z)+
{dz/dv\over (1+z)^2Y(z)d(0;z)}=
\left[{d\ \over dv}d(0;z)\right]\chi(z_i;z)+{1\over d(0;z)}.
\label{eq:firstdervf}
\end{equation}
The second derivative of $f(z)$ with respect to $v$ is also given by
\begin{equation}
{d^2\ \over dv^2}f(z)=\left[{d^2\ \over dv^2}d(0;z)\right]\chi(z_i;z)
=-{4\pi G\over H_0^2}\bar{\alpha}\bar{\rho}(z)(1+z)^2f(z).
\label{eq:secderivf}
\end{equation}
Therefore we find that $f(z)$ is a solution of the Dyer-Roeder equation 
(\ref{eq:DREQ}). \par
Next we investigate what initial conditions $f(z)$ satisfies. 
It is evident from equations (\ref{eq:chiddd}) and (\ref{eq:firstdervf}) 
that $f(z_i)=0$ and
\begin{equation}
 {d\ \over dv}f(z)\Bigr|_{z=z_i}
={dz/dv|_{z=z_i}\over (1+z_i)^2Y(z_i)d(0;z_i)}
={1\over d(0;z_i)}.
\label{eq:initf}
\end{equation}
From equations (\ref{eq:secderivf}) and (\ref{eq:initf}), we find that 
$(1+z_i)d(0;z_i)f(z)$ is equivalent to $d(z_i;z)$.
\par
The proof of equation (\ref{eq:chiddd}) is thus completed. 
Functions similar to the $\chi$-function are defined in different forms
by several authors \citep{NH84,PS88} for their own
purposes. \citeauthor{NH84} defined a distance as an optical distance,
which is equivalent to the inverse of the
$\chi$-function. \citeauthor{PS88} defined the function as an
independent solution of the usual Mattig distance which satisfies the
Raychaudhuri equation of the null geodesic in the homogeneous universe
model.
\section{Some Proofs}
\subsection{Equation (13) is a solution of equation (11)}
In this appendix, we prove that in the continuous limit, the angular 
position $\bmath{\theta}(z)$ of images in the lensing plane at redshift
$z$ satisfies equation (\ref{eq:cmlp}). First of all, we differentiate
equation (\ref{eq:formaltheta}) with respect to redshift $z$ and then
obtain 
\begin{equation}
{d\ \over dz}\bmath{\theta}(z)=-{3\over2}\Omega_0
\int_0^zd\zeta
{(1+\zeta)^2d(0;\zeta)\over Y(\zeta)}\left[{d\ \over 
dz}\left\{{d(\zeta;z)\over 
d(0;z)}\right\}\right]\frac{1}{\pi}\iint_{\cal D}
d^2\bmath{\theta}'\Delta_{\bar{\alpha}}(\bmath{\theta}',Z(\zeta))
{\bmath{\theta}(\zeta)-\bmath{\theta}'
  \over|\bmath{\theta}(\zeta)-\bmath{\theta}'|^2}.
\label{eq:org}
\end{equation}
After multiplying  $dz/d\chi(=-(1+z)^2d^2(0;z)Y(z))$ to  both sides of
this equation, we differentiate it with respect to $z$, again.
\begin{eqnarray}
{d\ \over dz}\left[{dz\over d\chi}{d\ \over dz}\bmath{\theta}(z)\right]&=&
{3\over2}\Omega_0
(1+z)^4d^3(0;z)
\left[{d\ \over dz}\left\{{d(\zeta;z)\over d(0;z)}\right\}\right]_{\zeta=z}\nonumber\\
&&\hspace{30mm}\times
\frac{1}{\pi}
\iint_{\cal D}d^2\bmath{\theta}'
\Delta_{\bar{\alpha}}(\bmath{\theta}',Z(z))
{\bmath{\theta}(z)-\bmath{\theta}'\over|\bmath{\theta}(z)-\bmath{\theta}'|^2}
\nonumber\\
&&+{3\over2}\Omega_0
\int_0^zd\zeta
{(1+\zeta)^2d(0;\zeta)\over Y(\zeta)}
{d\ \over dz}\left[(1+z)^2d^2(0;z)Y(z){d\ \over dz}
\left\{{d(\zeta;z)\over d(0;z)}\right\}\right]\nonumber\\
&&\hspace{40mm}\times
\frac{1}{\pi}\iint_{\cal D}d^2\bmath{\theta}'
\Delta_{\bar{\alpha}}(\bmath{\theta}',Z(\zeta))
{\bmath{\theta}(\zeta)-\bmath{\theta}'
  \over|\bmath{\theta}(\zeta)-\bmath{\theta}'|^2}.
\label{eq:2ndD}
\end{eqnarray}
Here, we apply the initial conditions of the Dyer-Roeder angular 
diameter distance, (\ref{eq:Hubblelaw}), to the first term of the right
hand side of equation (\ref{eq:2ndD}) to obtain
\begin{equation}
\left[{d\ \over dz}\left\{{d(\zeta;z)\over d(0;z)}\right\}\right]_{\zeta=z}
={1\over(1+z)Y(z)d(0;z)}.
\label{eq:RSH1}
\end{equation}
In the second term of the right hand side of equation (\ref{eq:2ndD}), 
equation (\ref{eq:vz}) provides that
\begin{eqnarray}
&&{d\ \over dz}\left\{(1+z)^2d^2(0;z)Y(z){d\ \over dz}
\left[{d(\zeta;z)\over d(0;z)}\right]\right\}\nonumber\\&&=
{1\over(1+z)^2Y(z)}{d\ \over dv}
\left\{
 d^2(0;z){d\ \over dv}\left[
		       {d(\zeta;z)\over d(0;z)}
		      \right]
\right\}\nonumber\\
 &&={1\over(1+z)^2Y(z)}\left[
d(0;z){d^2\ \over dv^2}d(\zeta;z)-d(\zeta;z){d^2\ \over dv^2}d(0;z)
\right]
\end{eqnarray}
vanishes because both $d(0;z)$ and $d(\zeta;z)$ are solutions of 
the same equation (\ref{eq:DREQ}). Finally we find that equation 
(\ref{eq:2ndD}) is rewritten as
\begin{equation}
{d\ \over dz}\left[{dz\over d\chi}{d\ \over dz}\bmath{\theta}(z)\right]=
{3\over2}\Omega_0
{(1+z)^3d^2(0;z)\over Y(z)}
\frac{1}{\pi}\iint_{\cal D}
d^2\bmath{\theta}'\Delta_{\bar{\alpha}}(\bmath{\theta}',Z(z))
{\bmath{\theta}(z)-\bmath{\theta}'
 \over|\bmath{\theta}(z)-\bmath{\theta}'|^2}.
\end{equation}
Similarly we can prove that equation (\ref{eq:sMag}) satisfies equation
(\ref{eq:cMag}). 
\subsection{About equation (17)}
In this appendix, we prove that if arbitrary function $g(z)$ has a finite 
value at $z=0$, 
\begin{equation}
\int_0^zd\zeta{(1+\zeta)^2d(0;\zeta)d(\zeta;z)\over d(0;z)Y(z)}g(\zeta)=
\int_{\chi(0)}^{\chi(z)}d\chi_1\int_{\chi(0)}^{\chi(\zeta_1)}d\chi_2
(1+\zeta_2)^5d^4(0;\zeta_2)g(\zeta_2),
\label{eq:prooftargetC}
\end{equation}
where $\chi_{1,2}=\chi(\zeta_{1,2})$ defined in the Appendix A3. This 
relation is used in the text when equation (\ref{eq:SolMag}) is derived 
from equation (\ref{eq:sMag}).\par
To begin with, we replace $d(\zeta;z) $in the left hand side of
(\ref{eq:prooftargetC}) by the $\chi$-function by using relation 
(\ref{eq:chiddd}):
\begin{eqnarray}
&&\int_0^zd\zeta{(1+\zeta)^2d(0;\zeta)d(\zeta;z)\over 
d(0;z)Y(\zeta)}g(\zeta)\nonumber\\
&=&\int_0^zd\zeta(1+\zeta)^3d^2(0;\zeta)
{\chi(\zeta)-\chi(z)\over Y(\zeta)}g(\zeta)\nonumber\\
&=&\int_0^zd\zeta\chi(\zeta){(1+\zeta)^3d^2(0;\zeta)\over Y(\zeta)}g(\zeta)
-\chi(z)\int_0^zd\zeta
{(1+\zeta)^3d^2(0;\zeta)\over Y(\zeta)}g(\zeta)\nonumber\\
&=&-\int_{\chi(0)}^{\chi(z)}d\chi_1\chi_1(1+\zeta_1)^5d^4(0;\zeta_1)g(\zeta_1)
+\chi(z)\int_{\chi(0)}^{\chi(z)}d\chi_1(1+\zeta_1)^5d^4(0;\zeta_1)g(\zeta_1).
\label{eq:LHSC2}
\end{eqnarray}
In equation (\ref{eq:LHSC2}), we introduce a function
\begin{displaymath}
h(\chi)\equiv\int_{\chi(0)}^\chi d\chi_1(1+\zeta_1)^5d^4(0;\zeta_1)g(\zeta_1),
\end{displaymath}
and rewrite equation (\ref{eq:LHSC2}) as
\begin{eqnarray}
\int_0^zd\zeta{(1+\zeta)^2d(0;\zeta)d(\zeta;z)\over 
d(0;z)Y(\zeta)}g(\zeta)&=&\chi h(\chi)-\int_{\chi(0)}^{\chi(z)}d\chi_1 
\chi_1{d\ \over d\chi_1}h(\chi_1)\nonumber\\
&=&\chi(0)h(\chi(0))
+\int_{\chi(0)}^{\chi(z)}d\chi_1h(\chi_1),
\end{eqnarray}
where we define $\chi(0)h(\chi(0))$ as a value of 
$\chi(\epsilon)h(\chi(\epsilon))$ in the limit of $\epsilon\to0$. 
The behaviors of $\chi(\epsilon)$ and $h(\chi(\epsilon))$ for small 
$\epsilon$ are $\chi(\epsilon)\sim O(1/\epsilon)$ and 
$h(\chi(\epsilon))\sim\epsilon^3g(\epsilon)$.
We thus find that $\chi(\epsilon)h(\chi(\epsilon))\sim 
\epsilon^2g(\epsilon)\to0$ if $g(0)$ is finite. Finally we have 
completed the proof of equation (\ref{eq:prooftargetC}).

\end{document}